\documentclass[%
 reprint,
 amsmath,amssymb,
 aps,
]{revtex4}

\usepackage{graphicx}
\usepackage{dcolumn}
\usepackage{bm}
\usepackage{subfigure}

\begin{document}

\preprint{APS/123-QED}

\title{Stress-energy tensor of quantized scalar fields in a zero-tidal wormhole}

\author{Shun Jiang}
\email{shunjiang@mail.bnu.edu.cn}
\author{Jie Jiang}
\email{Corresponding author. jiejiang@mail.bnu.edu.cn}
\affiliation{School of Physics and Optoelectronics, South China University of Technology, Guangzhou 510641, China\\}
\affiliation{Faculty of Arts and Sciences, Beijing Normal University, Zhuhai 519087, China}
\date{\today}

\begin{abstract}
To create a static traversable wormhole, exotic matter that meets the Morris-Thorne conditions is required. It is well known that the expectation value of the vacuum stress-energy tensor can violate the null energy condition, and thus has long been considered the best candidate for exotic matter. In this paper, we investigate whether the renormalized‌ stress-energy tensor of a non-minimally coupled massive scalar field in a zero-tidal wormhole can satisfy the Morris-Thorne conditions. Within the Hadamard renormalization framework, we calculate the renormalized stress-energy tensor using the pragmatic mode-sum regularization method recently established by Levi and Ori.
By varying the scalar field mass $m_0$ and coupling constant $\xi$, we find that there are three disconnected regions in this two-dimensional parameter space that satisfy the Morris-Thorne conditions. We identify two intervals in the scalar field mass $m_0$, within which the Morris-Thorne conditions cannot be satisfied irrespective of the value of the coupling constant $\xi$. This establishes two mass exclusion regions that constitute a ‌no-go regime‌ for the construction of traversable wormholes.
\end{abstract}


\maketitle

\section{INTRODUCTION}

The concept of traversable wormholes was brought into the physical domain by Morris and Thorne \cite{Morris:1988cz} in 1988. To maintain the structure of a wormhole, they found that the matter at the wormhole throat needs to satisfy the Morris-Thorne conditions, and this will lead to a violation of energy conditions. Due to this exotic behavior, they referred to the matter at the wormhole throat as exotic matter. It is well known that the quantum stress-energy tensor can violate energy conditions. Accordingly, they proposed that it could serve as a candidate for exotic matter.

‌To investigate whether a quantum scalar field can serve as exotic matter, analytical approximations of the quantum the stress-energy tensor has been employed to examine whether the the Morris-Thorne conditions conditions can be satisfied \cite{Matyjasek:2020cmi,Popov:2005qy,Carlson:2010yw,Kocuper:2017aap,Taylor:1996yu,Popov:2001kk,Khusnutdinov:2003ii}. Calculating the exact quantum vacuum energy-momentum tensor has always been a difficult task. To avoid this difficulty, a short-throat flat-space wormhole model has been studied in \cite{Bezerra:2010ix}. This model consists of two identical copies of Minkowski space, each with spherical regions excised, whose boundaries are to be identified. Thus the spacetime is everywhere flat except a two-dimensional singular spherical surface. Within this model setup, the vacuum stress-energy tensor has been obtained \cite{Bezerra:2010ix}. It represents an infinitesimally thin exotic matter layer at the wormhole throat, indicating a discontinuous distribution in spacetime. To gain a deeper understanding of ‌the behavior of the quantum stress-energy tensor‌ at the wormhole throat, it is natural to calculate the exact renormalized vacuum stress-energy tensor in more realistic wormhole models. 

The naive computation of the quantum stress-energy tensor leads to a divergent result. Thus, renormalization must be considered, which constitutes the primary challenge in computing the quantum stress-energy tensor. Using point-splitting method \cite{Schwinger:1951nm,DeWitt:1964mxt}, the expression for the divergent terms of the energy-momentum tensor was obtained in \cite{Christensen:1976vb}. In 1984, Candelas and Howard \cite{Candelas:1984pg} developed a method to calculate the expectation value of $\phi^2$. They applied this method to compute the expectation value of $\phi^2$ in Schwarzschild spacetime \cite{Candelas:1984pg}. Subsequently, Howard \cite{Howard:1984ttx} used this method to calculate the expectation value of the stress-energy tensor in Schwarzschild spacetime. This approach involves using a Wick rotation to convert the Lorentzian metric into a Euclidean one. Recently, Levi and Ori \cite{Levi:2015eea,Levi:2016paz} proposed the pragmatic mode-sum regularization method for computing the renormalized stress-energy tensor. This approach allows direct calculations in Lorentzian backgrounds. When applying the point-splitting technique in the two aforementioned regularization methods, it is necessary to subtract the corresponding counter terms to remove the  divergences. In \cite{Christensen:1976vb}, the explicit expression of the counter term is provided. However, it takes the form of a power series in terms of $1/m_0^2$, which is inconvenient for explicit calculations. The Hadamard renormalization approach directly computes the divergent terms by exploiting the singular structure of the two-point correlation function, thereby avoiding the computational complexity arising from series expansions. The Hadamard renormalization for a quantized scalar field has been developed in \cite{Decanini:2005eg}. By employing the bitensor expansion technique outlined in \cite{Christensen:1976vb} and the geometric expressions given in reference \cite{Decanini:2005eg}, one can in principle derive the counter terms required for Hadamard renormalization.

To our best knowledge, the exact vacuum quantum stress-energy tensor for wormholes has so far been calculated only under the  short-throat flat-space approximation \cite{Bezerra:2010ix}. To investigate whether it can sustain the wormhole throat, we must transition to more realistic wormhole models. In this paper, we study the renormalized stress-energy tensor of a massive scalar field with non-minimal coupling at the throat of a zero-tidal wormhole which is the simplest traversable wormhole with zero radial tides. Within the Hadamard renormalization framework \cite{Decanini:2005eg}, we employ the mode-sum regularization method developed by Levi and Ori \cite{Levi:2015eea,Levi:2016paz} to compute the renormalized stress-energy tensor. We explore the parameter space of the scalar field’s mass and  coupling constant to identify regions where the Morris-Thorne conditions are fulfilled.

This paper is organized as follows. In Section \ref{section2}, we briefly review the Morris-Thorne conditions and the zero-tidal wormholes. In Section \ref{section3}, we quantize scalar fields in a zero-tidal wormhole spacetime, present the singular structure of its two-point correlation function, which is known as the Hadamard condition, and discuss the implementation of Hadamard renormalization. In Section \ref{section4}, we numerically calculate the renormalized stress-energy tensor and examine whether it satisfies the Morris-Thorne conditions at the wormhole throat. In Section \ref{section5}, we discuss the results.

\section{ZERO-TIDAL WORMHOLES AND THE MORRIS-THORNE CONDITIONS}\label{section2}

In this section, we first briefly review the general traversable wormhole model, as well as the energy conditions required for traversable wormholes, namely the Morris-Thorne conditions. Then, we introduce the zero-tidal wormhole model, the simplest traversable wormhole with zero radial tides we will consider in this paper.

The metric of the static and  spherically symmetric wormhole can be written as \cite{Morris:1988cz}
\begin{eqnarray}
	ds^2=-e^{2\Phi(r)}dt^2+\left(1-\frac{b(r)}{r}\right)^{-1}dr^2+r^2d\Omega^2. \label{general metric form}
\end{eqnarray}
Here $\Phi(r)$ is the redshift function and $b(r)$ is the shape function. To achieve the traversable wormhole geometry, $\Phi(r)$ and $b(r)$ are required to satisfy the following conditions \cite{Morris:1988cz} :

(A1) There exist an $r_0>0$ such that $b(r_0)=r_0$ and $1-\frac{b(r)}{r}\geq0$ for $r\geq r_0$.

(A2) $\frac{d^2r}{dr_*^2}>0$ near $r_0$, where $r_*(r)=\pm\int_{r_0}^{r}\left(1-\frac{b(r)}{r}\right)^{-\frac{1}{2}}dr$ is the proper radial distance.

(A3) $\Phi(r)$ is everywhere finite.

Item (A1) and (A2) imply the existence of a wormhole throat at $r=r_0$. Item (A3) ensures the absence of a horizon, thereby rendering the wormhole traversable. 

Now, let us consider what kind of stress-energy tensor is required for traversable wormholes. Let $\{e^a_{u}\}$ be an orthonormal tetrad corresponding to the static observer, which can be written as
\begin{eqnarray}
	&&e^a_{0}=e^{-\Phi(r)}\left(\frac{\partial}{\partial t}\right)^a,\quad e^a_1=\left(1-\frac{b(r)}{r}\right)^{\frac{1}{2}}\left(\frac{\partial}{\partial r}\right)^a,\nonumber\\
	&&e^a_2=\frac{1}{r}\left(\frac{\partial}{\partial \theta}\right)^a,\quad e^a_3=\frac{1}{r\sin\theta}\left(\frac{\partial}{\partial \phi}\right)^a.
\end{eqnarray}
Using this set of frames, the Einstein field equations can be written as
\begin{eqnarray}
	\rho=T_{00}=\frac{1}{8\pi r^2}\frac{d}{dr}b(r),\\
	\tau=-T_{11}=\frac{1}{8\pi}\left[\frac{b}{r^3}-\frac{2}{r}\left(1-\frac{b}{r}\right)\frac{d}{dr}\Phi(r)\right],\\
	p=T_{22}=T_{33}=\frac{r}{2}\left[(\rho-\tau)\frac{d}{dr}\Phi(r)-\frac{d}{dr}\tau\right]-\tau.
\end{eqnarray}
Using items (A1)–(A3), the conditions for making a wormhole traversable can be expressed as
\begin{eqnarray}
	\tau_{0}>0,\label{MT1}\\
	\eta_0=\tau_{0}-\rho_{0}>0\label{MT2},
\end{eqnarray}
where $\tau_{0}=\tau|_{r=r_0}$ and  $\rho_{0}=\rho|_{r=r_0}$. Conditions (\ref{MT1}) and (\ref{MT2}) are called the Morris-Thorne conditions. The difficulty is that the Morris-Thorne conditions are not easily satisfied. From condition (\ref{MT1}), we can see that $\tau>0$ implies a tension exists at the wormhole throat, rather than a pressure as would correspond to a perfect fluid for $\tau<0$. Furthermore, condition (\ref{MT2}) leads to the violation of energy conditions by the matter at the wormhole throat. Therefore, matter satisfying conditions (\ref{MT1}) and (\ref{MT2}) is referred to as exotic matter. It is well known that the vacuum energy-momentum tensor in quantum field theory can violate energy conditions. Thus, quantum vacuum fluctuations might maintain the geometric structure of a wormhole. 

In this paper, we consider the zero-tidal wormhole which is the simplest traversable wormhole with the zero radial tides. By choosing $\Phi(r)=0$ and $b(r)=b_0$ in (\ref{general metric form}), the line element of the zero-tidal wormhole can be written as
\begin{eqnarray}
ds^2=-dt^2+\left(1-\frac{b_0}{r}\right)^{-1}dr^2+r^2d\Omega^2.
\end{eqnarray}

To explore the parameter space of spacetime and scalar fields, we will consider the Morris-Thorne conditions within the dimensionless $m_0b_0-\xi$ parameter space. To simplify the notation, we set
$b_0=1$, allowing us to identify $m_0b_0$ with $m_0$.

\section{Quantum Field Theory in Wormhole SpaceTime}\label{section3}

In this section, we start by quantizing the Klein-Gordon scalar field in the zero-tidal wormhole spacetime. We then introduce the Hadamard condition, which all physically reasonable quantum states must satisfy, as it specifies the short-distance singular behavior of the two-point correlation function. Therefore, to obtain the expectation value of the renormalized stress-energy tensor, we introduce the Hadamard renormalization procedure at the end of this section.

We begin by reviewing the classical Klein-Gordon scalar field. The associated action is given by
\begin{eqnarray}
	S=-\frac{1}{2}\int d^4x\sqrt{-g}\left(g^{ab}\nabla_a\phi\nabla_b\phi+m_0^2\phi^2+\xi R\phi^2\right),
\end{eqnarray}
where $m_0$ is the mass of the scalar field and $\xi$ is the coupling constant of the scalar field. The equation for a non-minimally coupled massive scalar field can be written as
\begin{eqnarray}
	\left(\nabla_a\nabla^a-m_0^2-\xi R\right)\phi=0.
\end{eqnarray}
The stress-energy tensor is given by
\begin{eqnarray}
	T_{uv}=\left(1-2\xi\right)\phi_{;u}\phi_{;v}+\left(2\xi-\frac{1}{2}\right)g_{uv}g^{\rho\sigma}\phi_{;\rho}\phi_{;\sigma}-2\xi\phi\phi_{;uv}\nonumber\\
	+2\xi g_{uv}\phi\nabla_a\nabla^a\phi+\xi\left(R_{uv}-\frac{1}{2}Rg_{uv}\right)\phi^2-\frac{1}{2}g_{uv}m_0^2\phi^2. \label{classical stress energy tensor}
\end{eqnarray}

Now, we quantize the Klein-Gordon scalar field. Using spherical symmetry, the mode functions can be written as
\begin{eqnarray}
	\psi_{\omega lm}(t,r,\theta,\varphi)=\frac{1}{\sqrt{4\pi|\omega|}}e^{-i\omega t}Y_{l m}(\theta,\varphi)\frac{R_{\omega l}(r)}{r}.
\end{eqnarray}
The radial funciton $R_{\omega l}(r)$ satisfies the following equation
\begin{eqnarray}
	\frac{d^2R_{\omega l}}{dr_{*}^2}+\left(\omega^2-V_{\text{eff}}\right)R_{\omega l}=0,\label{radial equation}
\end{eqnarray}
where the effective potential is given by
\begin{eqnarray}
V_{\text{eff}}(r)=f(r)\left(m_0^2+\xi R+\frac{l(l+1)}{r^2}+\frac{1}{r\sqrt{f(r)h(r)}}\frac{d}{dr}\left(\sqrt{\frac{f(r)}{h(r)}}\right)\right).
\end{eqnarray}
Here we take
\begin{eqnarray}
	f(r)=1,\quad h(r)=\left(1-\frac{b_0}{r}\right)^{-1}.
\end{eqnarray}
Following \cite{DeWitt:1975ys}, we choose two linearly independent basis functions for equation (\ref{radial equation}), denoted $R^{L}_{\omega l}$ and $R^{R}_{\omega l}$. The boundary conditions for $R^{L}_{\omega l}$ and $R^{R}_{\omega l}$ are given by

\begin{eqnarray}
R_{\omega l}^{L}=\left\{\begin{array}{lr}
e^{i\sqrt{\omega^2-m_0^2}r_{*}}+\mathcal{R}^{L}_{\omega l}e^{-i\sqrt{\omega^2-m_0^2} r_{*}}, & r_{*}\rightarrow-\infty,\\\
\mathcal{T}^{L}_{\omega l}e^{i\sqrt{\omega^2-m_0^2} r_{*}},     & r_{*}\rightarrow\infty,\\
\end{array}\right.\quad
R_{\omega l}^{R}=\left\{\begin{array}{lr}
e^{-i\sqrt{\omega^2-m_0^2} r_{*}}+\mathcal{R}^{R}_{\omega l}e^{i\sqrt{\omega^2-m_0^2} r_{*}}, & r_{*}\rightarrow\infty,\\\
\mathcal{T}^{R}_{\omega l}e^{-i\sqrt{\omega^2-m_0^2} r_{*}},     & r_{*}\rightarrow-\infty.\\
\end{array}\right.\label{boundary condition1}
\end{eqnarray}
Thus, we have two linearly independent mode functions $\psi^{L}_{\omega lm}$ and $\psi^{R}_{\omega lm}$ which are given by
\begin{eqnarray}
\psi^{L/R}_{\omega lm}(t,r,\theta,\varphi)=\frac{1}{\sqrt{4\pi|\omega|}}e^{-i\omega t}Y_{l m}(\theta,\varphi)\frac{R^{L/R}_{\omega l}(r)}{r}. \label{mode funtion}
\end{eqnarray}
The field operator can then be conveniently expressed as
\begin{eqnarray}
	\hat{\phi}(x)=\int_{0}^{\infty}dk\sum_{l=0}^{\infty}\sum_{m=-l}^{m=l}\left(a_{\omega(k) lm}^{L}\psi^{L}_{\omega(k) lm}(x)+a_{\omega(k) lm}^{R}\psi^{R}_{\omega(k) lm}(x)+a_{\omega(k) lm}^{L\dagger}\psi^{L*}_{\omega(k) lm}(x)+a_{\omega(k) lm}^{R\dagger}\psi^{R*}_{\omega(k) lm}(x)\right).
\end{eqnarray}
Here $\omega(k)=\sqrt{m_0^2+k^2}$. 

The vacuum state $|0\rangle$, corresponding to the annihilation operators $a^{L}_{\omega(k)lm}$ and  $a^{R}_{\omega(k)lm}$, is defined by
\begin{eqnarray}
	a^{L}_{\omega(k)lm}|0\rangle=0,\quad a^{R}_{\omega(k)lm}|0\rangle=0.
\end{eqnarray}
Then, we may calculate vacuum expectation value of $\hat{\phi}^2$, and its expression is
\begin{eqnarray}
	\langle0|\hat{\phi}^2(x)|0\rangle=\int_{0}^{\infty}dk\sum_{l=0}^{\infty}\sum_{m=-l}^{m=l}\left(\psi^{L*}_{\omega(k) lm}(x)\psi^{L}_{\omega(k) lm}(x)+\psi^{R*}_{\omega(k) lm}(x)\psi^{R}_{\omega(k) lm}(x)\right)\label{mode sum}
\end{eqnarray}

However, one may find that this expression diverges. This is because physically reasonable states need to satisfy the Hadamard condition \cite{Radzikowski:1996pa,Hollands:2019whz,Kay:1988mu}, which describes the short-distance singular behavior of the two-point correlation function. 
It is useful to introduce a few concepts before presenting the Hadamard condition. The convex normal neighborhood is a globally hyperbolic sub-spacetime $U\subset M$ such that any $x,x'\in U$ can be connected by a unique geodesic within $U$. For any $x,x'\in U$, we can define the signed squared geodesic distance $\sigma(x,x')$ uniquely \cite{Radzikowski:1996pa}. Since $U$ is a time oriented sub-spacetime, we can define a time function $T(x)$ on $U$. The Hadamard condition can be defined as follow \cite{Kay:1988mu}. For any $x,x'\in U$, the two point correlation function $\omega_{2}(x,x')$ has the general form
\begin{eqnarray}
\omega_2(x,x')=\lim_{\epsilon\rightarrow0^{+}}\frac{1}{4\pi^2}\left(\frac{\Delta^{\frac{1}{2}}(x,x')}{\sigma+i\epsilon T}+V(x,x')\text{log}(\sigma+i\epsilon T)+W_{\omega}(x,x')\right), \label{hadamard condition}
\end{eqnarray}
where $\Delta^{1/2}$ is the square root of Van Vleck-Morette determinant, $V$ can be obtained by Hadamard recursion relation \cite{DeWitt:1960fc}, $T=T(x)-T(x')$, $W_{\omega}$ is a state dependent smooth function on $U\times U$ and $\epsilon\rightarrow0$ in the distribution sense. 

It should be noted that the first two terms on the right-hand side of the equation (\ref{hadamard condition}) exhibit singular behavior when $x\rightarrow x'$. Due to this singular behavior, the expression (\ref{mode sum}) diverges and requires renormalization. Fortunately, $\Delta^{1/2}$ and $V$ are determined uniquely by the local geometry, which makes renormalization possible.

In this paper, we use the Hadamard renormalization scheme \cite{Decanini:2005eg}. First, let's recall how to obtain the renormalized expectation values of $\hat{\phi}^2$ within the Hadamard framework. For any two-point correlation function $\omega_{2}(x,x')$ given in (\ref{hadamard condition}), the corresponding renormalized vacuum expectation value of $\hat{\phi}^2$ is formally given by
\begin{eqnarray}
	\omega_{2,\text{rem}}(x)=\lim_{x'\rightarrow x}\frac{1}{2}\left(\omega_{2}(x,x')+\omega_{2}(x',x)\right)-C(x,x')=\lim_{x'\rightarrow x}\frac{1}{4\pi^2}W_{\omega}(x,x'),
\end{eqnarray}
where the symbol $\omega_{2,\text{rem}}$ represents the corresponding renormalized vacuum expectation value of $\hat{\phi}^2$ and $C(x,x')$ is given by
\begin{eqnarray}
	C(x,x')=\frac{1}{4\pi^2}\left(\frac{\Delta^{\frac{1}{2}}(x,x')}{\sigma}+V(x,x')\text{log}|\sigma|\right).
\end{eqnarray}
Here, for convenience, we use the symmetric two point function to eliminate the $\epsilon$ part in (\ref{hadamard condition}).

Similarly, according to the expression of the classical energy-momentum tensor (\ref{classical stress energy tensor}), the renormalized vacuum energy-momentum tensor $T_{uv,\text{rem}}$ can be written as
\begin{eqnarray}
	T_{uv,\text{rem}}(x)&&=\lim_{x'\rightarrow x}\tilde{T}_{uv}(x,x')\left[\frac{1}{2}\left(\omega_{2}(x,x')+\omega_{2}(x',x)\right)-C(x,x')\right]+\frac{v_1}{4\pi^2}g_{uv}\nonumber\\
	&&=\lim_{x'\rightarrow x}\frac{1}{4\pi^2}\tilde{T}_{uv}(x,x')W_{\omega}(x,x')+\frac{v_1}{4\pi^2}g_{uv}, \label{renormal energy tensor}
\end{eqnarray}
where the second term on the right-hand side corresponds to the trace anomaly \cite{Wald:1978pj} and
the $\tilde{T}_{uv}(x,x')$ is given by \cite{Decanini:2005eg}
\begin{eqnarray}
	\tilde{T}_{uv}(x,x')=\left(1-2\xi\right)g_v{ }^{v'}\nabla_u\nabla_{v'}+\left(2\xi-\frac{1}{2}\right)g_{uv}g^{\rho\sigma'}\nabla_{\rho}\nabla_{\sigma'}-2\xi g_{u}{ }^{u'}g_{v}{ }^{v'}\nabla_{u'}\nabla_{v'}\nonumber\\
    +2\xi g_{uv}\nabla_\rho\nabla^\rho+\xi\left(R_{uv}-\frac{1}{2}Rg_{uv}\right)-\frac{1}{2}m_0^2g_{uv},
\end{eqnarray}
where $g_{u}{ }^{u'}$ is the bivector of parallel displacement \cite{Christensen:1976vb}.

In practice, we know that expressions are represented in the form of mode functions, such as (\ref{mode sum}), rather than in the form presented in (\ref{hadamard condition}). Therefore, we cannot obtain the analytical expression for $W_{\omega}(x,x')$ and it is convenient to utilize the first line of (\ref{renormal energy tensor}) for the calculation. For convenience, we define the symmetrized two-point correlation function, which can be written as
\begin{eqnarray}
G(x,x')=\langle0|\left[\hat{\phi}(x),\hat{\phi}(x')\right]_{+}|0\rangle=\int_{0}^{\infty}dk\sum_{l=0}^{\infty}\sum_{m=-l}^{m=l}\left(\{\psi^{L*}_{\omega(k) lm}(x),\psi^{L}_{\omega(k) lm}(x')\}+\{\psi^{R*}_{\omega(k) lm}(x),\psi^{R}_{\omega(k) lm}(x')\}\right), \label{mode expression}
\end{eqnarray}
where
\begin{eqnarray}
	\left[\hat{\phi}(x),\hat{\phi}(x')\right]_{+}=\frac{1}{2}\left(\hat{\phi}(x)\hat{\phi}(x')+\hat{\phi}(x')\hat{\phi}(x)\right)
\end{eqnarray}
and
\begin{eqnarray}
	\{X(x),Y(x')\}=\frac{1}{2}\left(X(x)Y(x')+Y(x)X(x')\right).
\end{eqnarray}
Similarly, the derivative $\nabla_u\nabla_v'$ of the symmetrized two-point correlation function is given by
\begin{eqnarray}
G(x,x')_{;uv'}&&=\langle0|\left[\nabla_{u}\hat{\phi}(x),\nabla_{v'}\hat{\phi}(x')\right]_{+}|0\rangle
\nonumber\\&&=\int_{0}^{\infty}dk\sum_{l=0}^{\infty}\sum_{m=-l}^{m=l}\left(\{\psi^{L*}_{\omega(k) lm}(x),\psi^{L}_{\omega(k) lm}(x')\}_{;uv'}+\{\psi^{R*}_{\omega(k) lm}(x),\psi^{R}_{\omega(k) lm}(x')\}_{;uv'}\right). \label{mode expression d2}
\end{eqnarray}
Other derivatives can be obtained in a similar manner. Thus, the expression of $T_{uv,\text{rem}}$ can be rewritten as
\begin{eqnarray}
	T_{uv,\text{rem}}(x)=\lim_{x'\rightarrow x}\left[\left(1-2\xi\right)\left(g_{v}{ }^{v'}G(x,x')_{;uv'}-g_{v}{ }^{v'}C(x,x')_{;uv'}\right)\right.\nonumber\\
	\left.+\left(2\xi-\frac{1}{2}\right)g_{uv}\left(g^{\rho\sigma'}G(x,x')_{;\rho\sigma'}-g^{\rho\sigma'}C(x,x')_{;\rho\sigma'}\right)-2\xi\left(g_u{ }^{u'}g_v{ }^{v'}G(x,x')_{;u'v'}-g_u{ }^{u'}g_v{ }^{v'}C(x,x')_{;u'v'}\right)\right.\nonumber\\
	\left.+2\xi g_{uv}\left(G(x,x')_{;\rho}{ }^{;\rho}-C(x,x')_{;\rho}{ }^{;\rho}\right)+\xi\left(R_{uv}-\frac{1}{2}Rg_{uv}\right)\left(G(x,x')-C(x,x')\right)-\frac{1}{2}m_0^2g_{uv}\left(G(x,x')-C(x,x')\right)\right].\label{final renormalized energy tensor}
\end{eqnarray}
Here $G(x,x')$ and its derivative can be obtained by solving the radial equation (\ref{radial equation}) numerically, while $C(x,x')$ and its derivative can be obtained using the method developed in \cite{Christensen:1976vb,Decanini:2005eg}.

\section{The Renormalized Stress-Energy Tensor and Morris-Thorne Conditions}\label{section4}

In this section, we utilize the pragmatic mode-sum regularization method \cite{Levi:2015eea,Levi:2016paz} within the Hadamard renormalization framework \cite{Decanini:2005eg} to calculate the renormalized stress-energy tensor for a scalar field at the throat of the zero-tidal wormhole. We vary the mass $m_0$ and coupling coefficient $\xi$ of the scalar field to analyze whether the quantum vacuum stress-energy tensor satisfies the Morris-Thorne conditions.

The expression for the renormalized stress-energy tensor is given by (\ref{final renormalized energy tensor}). For computational convenience, we need to rewrite (\ref{final renormalized energy tensor}) in a more manageable form. Here are four types of terms in (\ref{final renormalized energy tensor}), which can be written as
\begin{align}
	&1.\lim_{x'\rightarrow x}\left(g_{v}{ }^{v'}G(x,x')_{;uv'}-g_{v}{ }^{v'}C(x,x')_{;uv'}\right),\nonumber\\
	&2.\lim_{x'\rightarrow x}\left(g_u{ }^{u'}g_v{ }^{v'}G(x,x')_{;u'v'}-g_u{ }^{u'}g_v{ }^{v'}C(x,x')_{;u'v'}\right),\nonumber\\	
	&3.\lim_{x'\rightarrow x}\left(G(x,x')_{;uv}-C(x,x')_{;uv}\right),\nonumber\\
	&4.\lim_{x'\rightarrow x}\left(G(x,x')-C(x,x')\right). \label{four terms}
\end{align}

Let's consider the first term of (\ref{four terms}), which can be expressed as
\begin{eqnarray}
	G_{1uv,\text{rem}}(x)=\lim_{x'\rightarrow x}\left(g_{v}{ }^{v'}G(x,x')_{;uv'}-C_{1uv}(x,x')\right)\label{G1},
\end{eqnarray}
where
\begin{eqnarray}
	C_{1uv}(x,x')=g_{v}{ }^{v'}C(x,x')_{;uv'}.
\end{eqnarray}
As pointed out in \cite{Levi:2016paz}, for the sake of computational convenience, we rewrite (\ref{G1}) as‌
\begin{eqnarray}
G_{1uv,\text{rem}}(x)=\lim_{x'\rightarrow x}\left(G_{1uv}(x,x')-(g^{-1})_{v}{ }^{\sigma}C_{1u\sigma}(x,x')\right)\label{finalG1},
\end{eqnarray}
Here, $G_{1uv}(x,x')$ means we use the coordinate system of point $x$ to take the covariant derivative with respect to both $x$ and $x'$ once each.

Similarly, the second term of  (\ref{four terms}) can be written as
\begin{eqnarray}
G_{2uv,\text{rem}}(x)=\lim_{x'\rightarrow x}\left(G_{2uv}(x,x')-(g^{-1})_u{ }^{\rho}(g^{-1})_v{ }^{\tau}C_{2\rho\tau}(x,x')\right)
\end{eqnarray}
and the third term term of  (\ref{four terms}) can be written as
\begin{eqnarray}
G_{0uv,\text{rem}}(x)=\lim_{x'\rightarrow x}\left(G_{0uv}(x,x')-C_{0uv}(x,x')\right)\label{finalG0},
\end{eqnarray}
where
\begin{eqnarray}
C_{2uv}(x,x')=g_{u}{ }^{u'}g_{v}{ }^{v'}C(x,x')_{;u'v'}
\end{eqnarray}
and
\begin{eqnarray}
C_{0uv}(x,x')=C(x,x')_{;uv}.
\end{eqnarray}
Here, $G_{0uv}(x,x')$ corresponds to taking the covariant derivative twice with respect to $x$ using the coordinate system of point $x$. Similarly, $G_{2uv}(x,x')$ corresponds to taking the covariant derivative twice with respect to $x'$ using the coordinate system of point $x$.

Now, returning to (\ref{G1}), since the spacetime is static, we separate the points in the $t$-direction. We choose
\begin{eqnarray}
	x=\left(t,r,\theta,\varphi\right),\quad x'=\left(t+\epsilon,r,\theta,\varphi\right).
\end{eqnarray}
Then, the limit $x'\rightarrow x$ corresponds to $\epsilon\rightarrow0$. Using (\ref{mode funtion}), we have
\begin{eqnarray}
	\psi^{L/R}_{\omega lm}(x')=\psi^{L/R}_{\omega lm}(x)e^{-i\omega\epsilon}.
\end{eqnarray}
Then, inserting (\ref{mode expression d2}) into (\ref{finalG1}), we get
\begin{eqnarray}
G_{1uv,\text{rem}}(x)=\lim_{\epsilon\rightarrow 0}\int_{0}^{\infty} F_{1uv}(x,k)\cos\left(\omega(k)\epsilon\right)dk-L_{1uv}(x,\epsilon), \label{G1rem}
\end{eqnarray}
where
\begin{eqnarray}
	F_{1uv}(x,k)=\sum_{l=0}^{\infty}\sum_{m=-l}^{m=l}\left(\{\psi^{L*}_{\omega(k) lm,u}(x),\psi^{L}_{\omega(k) lm,v}(x)\}+\{\psi^{R*}_{\omega(k) lm,u}(x),\psi^{R}_{\omega(k) lm,v}(x)\}\right) \label{w1}
\end{eqnarray}
and
\begin{eqnarray}
	L_{1uv}(x,\epsilon)=(g^{-1})_{v}{ }^{\sigma}C_{1u\sigma}(x,x').
\end{eqnarray}
The specific expression of $L_{1uv}(x,\epsilon)$ can be obtained through the method in \cite{Christensen:1976vb}, and its general form can be written as
\begin{eqnarray}
	L_{1uv}(x,\epsilon)=\frac{a_{1uv}(x)}{\epsilon^4}+\frac{b_{1uv}(x)}{\epsilon^2}+\frac{c_{1uv}(x)}{\epsilon}+d_{1uv}(x)\log(|\sigma|)+e_{1uv}(x)+O(\epsilon).
\end{eqnarray}
It is convenient to rewrite the expression for $L_{1uv}(x,\epsilon)$ in integral form. To achieve it, we can utilize the following identities \cite{Anderson:1994hg,Levi:2016paz}
\begin{eqnarray}
	\int_{0}^{\infty}\omega^3\cos(\omega\epsilon)d\omega&&=\frac{6}{\epsilon^4},\nonumber\\
	\int_{0}^{\infty}\omega\cos(\omega\epsilon)d\omega&&=-\frac{1}{\epsilon^2},\nonumber\\
	\int_{0}^{\infty}\log(\sigma)\cos(\omega\epsilon)d\omega&&=-\frac{\pi}{2\epsilon},\nonumber\\
	\int_{0}^{\infty}\frac{1}{\omega+\mu e^{-\gamma}}\cos(\omega\epsilon)d\omega&&=-\log(\mu\epsilon)+O(\epsilon),
\end{eqnarray}
where $\gamma$ is the Euler constant.
Then, we can rewrite (\ref{G1rem}) as
\begin{eqnarray}
G_{1uv,\text{rem}}(x)=\lim_{\epsilon\rightarrow 0}\int_{0}^{\infty} F_{1uv}(x,k)\cos\left(\omega(k)\epsilon\right)dk-\lim_{\epsilon\rightarrow 0}\int_{0}^{\infty}F_{1uv,\text{sing}}(x,\omega(k))\cos\left(\omega(k)\epsilon\right)d\omega-e_{1uv}(x)\nonumber\\
=\lim_{\epsilon\rightarrow 0}\int_{0}^{\infty} \left[F_{1uv}(x,k)-\frac{k}{\sqrt{k^2+m_0^2}}F_{1uv,\text{sing}}(x,\omega(k))\right]\cos(\omega(k)\epsilon)dk\nonumber\\
-\lim_{\epsilon\rightarrow 0}\int_{0}^{m}F_{1uv,\text{sing}}(x,\omega(k))\cos\left(\omega(k)\epsilon\right)d\omega-e_{1uv}(x),\label{g11}
\end{eqnarray}
where
\begin{eqnarray}
F_{1uv,\text{sing}}(x,\omega)=\frac{1}{6}a_{1uv}(x)\omega^3-b_{1uv}(x)\omega-\frac{2}{\pi}c_{1uv}(x)\log(\omega)-d_{1uv}(x)\frac{1}{\omega+e^{-\gamma}}.
\end{eqnarray}

The limit and integral order can be directly exchanged in the third line of (\ref{g11}), whereas the interchange in the second line of (\ref{g11}) is non-trivial. We need to employ the self-cancellation technique developed in \cite{Levi:2015eea,Levi:2016paz} to handle this integral. By treating this integral as a generalized integral and applying self-cancellation to it, we can take the limit. Thus, we have
\begin{eqnarray}
G_{1uv,\text{rem}}(x)=\int_{0}^{\infty} \left[F_{1uv}(x,k)-\frac{k}{\sqrt{k^2+m_0^2}}F_{1uv,\text{sing}}(x,\omega(k))\right]dk
-\int_{0}^{m_0}F_{1uv,\text{sing}}(x,\omega(k))d\omega-e_{1uv}(x).\label{y1}
\end{eqnarray}
This gives the final computational form for the first term of (\ref{four terms}). Similarly, the expressions for the second and third terms of (\ref{four terms}) can be written as
\begin{eqnarray}
G_{2uv,\text{rem}}(x)=\int_{0}^{\infty} \left[F_{2uv}(x,k)-\frac{k}{\sqrt{k^2+m_0^2}}F_{2uv,\text{sing}}(x,\omega(k))\right]dk
-\int_{0}^{m_0}F_{2uv,\text{sing}}(x,\omega(k))d\omega-e_{2uv}(x) \label{y2}
\end{eqnarray}
and
\begin{eqnarray}
G_{0uv,\text{rem}}(x)=\int_{0}^{\infty} \left[F_{0uv}(x,k)-\frac{k}{\sqrt{k^2+m_0^2}}F_{0uv,\text{sing}}(x,\omega(k))\right]dk
-\int_{0}^{m_0}F_{0uv,\text{sing}}(x,\omega(k))d\omega-e_{0uv}(x), \label{y3}
\end{eqnarray}
where
\begin{eqnarray}
F_{2uv}(x,k)=\sum_{l=0}^{\infty}\sum_{m=-l}^{m=l}\left(\{\psi^{L*}_{\omega(k) lm}(x),\psi^{L}_{\omega(k) lm;uv}(x)\}+\{\psi^{R*}_{\omega(k) lm}(x),\psi^{R}_{\omega(k) lm;uv}(x)\}\right) \label{w2}
\end{eqnarray}
and
\begin{eqnarray}
F_{0uv}(x,k)=\sum_{l=0}^{\infty}\sum_{m=-l}^{m=l}\left(\{\psi^{L*}_{\omega(k) lm;uv}(x),\psi^{L}_{\omega(k) lm}(x)\}+\{\psi^{R*}_{\omega(k) lm;uv}(x),\psi^{R}_{\omega(k) lm}(x)\}\right). \label{w3}
\end{eqnarray}
Now we consider the fourth term of (\ref{four terms}), which can be written as
\begin{eqnarray}
G_{\text{rem}}(x)&&=\lim_{x'\rightarrow x}\left(G(x,x')-C(x,x')\right)\nonumber\\&&=\int_{0}^{\infty} \left[F(x,k)-\frac{k}{\sqrt{k^2+m_0^2}}F_{\text{sing}}(x,\omega(k))\right]dk
-\int_{0}^{m}F_{\text{sing}}(x,\omega(k))d\omega-e(x), \label{y4}
\end{eqnarray}
where
\begin{eqnarray}
F(x,k)=\sum_{l=0}^{\infty}\sum_{m=-l}^{m=l}\left(|\psi^{L}_{\omega(k) lm}(x)|^2+|\psi^{R}_{\omega(k) lm}(x)|^2\right) \label{w4}
\end{eqnarray}
and
\begin{eqnarray}
	F_{\text{sing}}(x,\omega)=-b(x)\omega-\frac{2}{\pi}c(x)\log(\omega)-d(x)\frac{1}{\omega+e^{-\gamma}}
\end{eqnarray}
Using the notation defined above, the renormalized stress-energy tensor (\ref{final renormalized energy tensor}) can be written as
\begin{eqnarray}
	T_{uv,\text{rem}}(x)=\left(1-2\xi\right)G_{1uv,\text{rem}}(x)+\left(2\xi-\frac{1}{2}\right)g_{uv}g^{\rho\tau}G_{1\rho\tau,\text{rem}}(x)-2\xi G_{2uv,\text{rem}}(x)\nonumber\\
	+2\xi g_{uv}g^{\rho\tau}G_{0uv,\text{rem}}+\xi\left(R_{uv}-\frac{1}{2}Rg_{uv}\right)G_{\text{rem}}-\frac{1}{2}m_0^2g_{uv}G_{\text{rem}}. \label{calculate final}
\end{eqnarray}
The expressions (\ref{y1}), (\ref{y2}), (\ref{y3}) and (\ref{y4}) provide us with a convenient form for calculating the stress-energy tensor. The $F_{iuv}$ and $F$ can be obtained by numerically solving radial equation (\ref{radial equation}). The $F_{iuv,\text{sing}}$ and $F$ can be derived via the method outlined in \cite{Christensen:1976vb}, combined with the explicit geometric forms given in \cite{Decanini:2005eg}. After some calculations, we obtain
\begin{eqnarray}
	&&a_{0tt}=a_{2tt}=-a_{1tt}=-\frac{3}{2\pi^2},\quad b_{0tt}=b_{2tt}=-b_{1tt}=-\frac{m_0^2}{8\pi^2},\quad c_{0tt}=c_{2tt}=-c_{1tt}=0,\nonumber\\
	&&d_{0tt}=d_{2tt}=-d_{1tt}=-\frac{1+20m_0^4r^6}{640\pi^2r^6},\quad e_{0tt}=e_{2tt}=-e_{1tt}=\frac{(1+20m_0^4r^6)(-3+\log 2)}{1280\pi^2r^6},\nonumber\\
	&&a_{0r_*r_*}=a_{2r_*r_*}=-a_{1r_*r_*}=-\frac{1}{2\pi^2},\quad b_{0r_*r_*}=b_{2r_*r_*}=-b_{1r_*r_*}=\frac{1-3m_0^2r^3}{24\pi^2r^3},\quad\nonumber\\
	&&c_{0r_*r_*}=c_{2r_*r_*}=-c_{1r_*r_*}=0,\quad d_{0r_*r_*}=d_{2r_*r_*}=-d_{1r_*r_*}=\frac{1-40m_0^2r^3+60m_0^4r^6}{1920\pi^2r^6},\quad\nonumber\\
	&&e_{0r_*r_*}=e_{2r_*r_*}=-e_{1r_*r_*}=-\frac{-3+60m_0^4r^6(-1+\log 2)+\log 2-40m_0^2r^3\log 2}{3840\pi^2r^6},\nonumber\\
	&&a_{0\varphi\varphi}=a_{2\varphi\varphi}=-a_{1\varphi\varphi}=-\frac{r^2}{2\pi^2},\quad b_{0\varphi\varphi}=b_{2\varphi\varphi}=-b_{1\varphi\varphi}=-\frac{1+6m_0^2r^3}{48\pi^2r},\quad\nonumber\\ 
	&&c_{0\varphi\varphi}=c_{2\varphi\varphi}=-c_{1\varphi\varphi}=0,\quad
	d_{0\varphi\varphi}=d_{2\varphi\varphi}=-d_{1\varphi\varphi}=\frac{-1+10m_0^2r^3+30m_0^4r^6}{960\pi^2r^4},\quad\nonumber\\ &&e_{0\varphi\varphi}=e_{2\varphi\varphi}=-e_{1\varphi\varphi}=\frac{3-60m_0^4r^6(-1+\log 2)-20m_0^2r^3\log 2+\log 4}{3840\pi^2r^4},\nonumber\\
	&&b=-\frac{1}{4\pi^2},\quad c=0,\quad d=\frac{m_0^2}{8\pi^2},\quad e=-\frac{m_0^2\log 2}{16\pi^2}.
\end{eqnarray}
 
Now, we turn to the calculation of the renormalized stress-energy tensor (\ref{calculate final}) using the pragmatic mode-sum regularization method \cite{Levi:2015eea,Levi:2016paz}. To obtain the values of $G_{iuv,\text{rem}}$ and $G_{\text{rem}}$ defined in (\ref{y1}), (\ref{y2}), (\ref{y3}), and (\ref{y4}), we need to numerically compute the values of the $F_{iuv}$ and $F$ which are given in (\ref{w1}), (\ref{w2}), (\ref{w3}), and (\ref{w4}). Since the calculation processes for these terms are very similar, we will present the detailed procedure for $G_{1uv,\text{rem}}$ as an example. 

\begin{figure}[h]
	\begin{center}
		\includegraphics[scale=0.4]{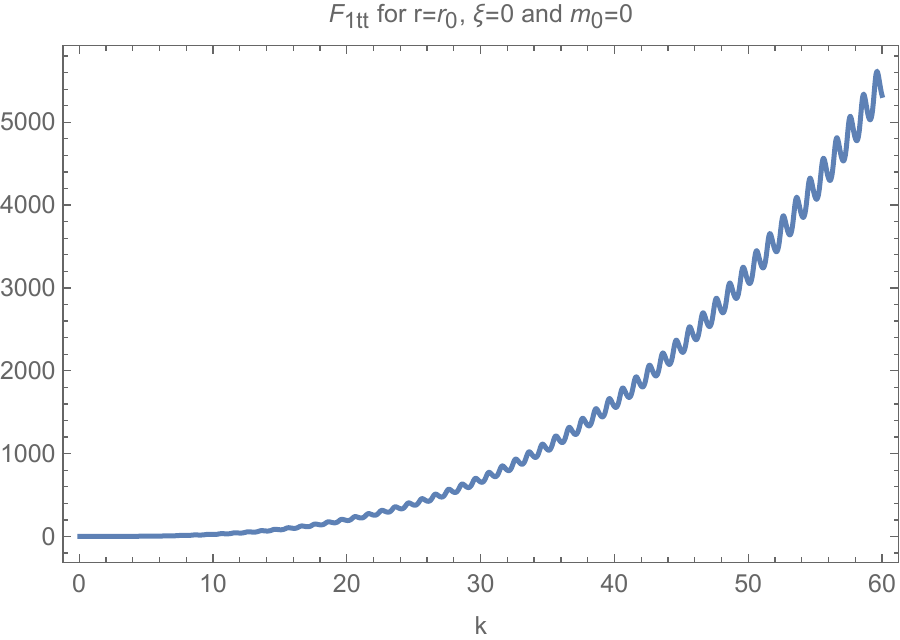}
	\end{center}
	\caption{The curve in this figure displays the value of $F_{\text{1tt}}$ for $r=r_0$, $\xi=0$ and $m_0=0$ in the zero-tidal wormhole.}
	\label{figure1}
\end{figure}

By numerically solving the radial equation  (\ref{radial equation}) with the given boundary conditions (\ref{boundary condition1}), we obtain $R^{L}_{\omega(k) lm}(x)$ and $R^{R}_{\omega(k) lm}(x)$ for $k$ between $0$ and $60$ with a
uniform step of $dk=1/20$. Using the relationship (\ref{mode funtion}) between $\psi^{L/R}_{\omega lm}$ and $R^{L/R}_{\omega lm}$, we obtain the explicit forms of $F_{1uv}$. We have
\begin{eqnarray}
	F_{\text{1tt}}(x,k)=\sum_{l=0}^{\infty}\sum_{m=-l}^{m=l}\frac{\omega}{4\pi r^2}\vert Y_{lm}(\theta,\varphi)\vert^2\left(\vert R^{L}_{\omega l}\vert^2+\vert R^{R}_{\omega l}\vert^2\right)\nonumber\\
	F_{\text{1}r_{*}r_{*}}=\sum_{l=0}^{\infty}\sum_{m=-l}^{m=l}\frac{1}{4\pi\omega}\vert Y_{lm}(\theta,\varphi)\vert^2\left(\left|\frac{d}{dr_*}\left(\frac{R^{L}_{\omega l}}{r}\right)\right|^2+\left|\frac{d}{dr_*}\left(\frac{R^{L}_{\omega l}}{r}\right)\right|^2\right)\nonumber\\
    F_{1\varphi\varphi}(x,k)=\sum_{l=0}^{\infty}\sum_{m=-l}^{m=l}\frac{1}{4\pi\omega r^2}\vert Y_{lm,\varphi}(\theta,\varphi)\vert^2\left(\vert R^{L}_{\omega l}\vert^2+\vert R^{R}_{\omega l}\vert^2\right).\label{F3}
\end{eqnarray}
The value of $F_{1\theta\theta}$ can be derived from $F_{1\varphi\varphi}$ by exploiting spherical symmetry. We can analytically sum over m by utilizing the following identities
\begin{eqnarray}
	\sum_{m=-l}^{m=l}\left|Y_{lm}(\theta,\varphi)\right|^2=\frac{2l+1}{4\pi},\nonumber\\
	\sum_{m=-l}^{m=l}\left|Y_{lm,\varphi}(\theta,\varphi)\right|^2=\frac{l(l+3l+2l^2)}{8\pi}.
\end{eqnarray}
For any fixed $\omega$, we sum over $l$ in (\ref{F3}) until the corresponding contribution falls below $10^{-12}$. The Fig.\ref{figure1} presents the numerical results of $F_{\text{1tt}}$ for $r=r_0$ with $m_0=0$. It indicates that the integral over k will diverge, and therefore renormalization must be considered. To better investigate the origin of the divergence, we define two quantities
\begin{eqnarray}
F_{1\text{ttreg}}(x,k)=F_{\text{1tt}}(x,k)-\frac{k}{\sqrt{k^2+m_0^2}}F_{\text{1tt,sing}}(x,\omega(k))
\end{eqnarray}
and
\begin{eqnarray}
H_{\text{1tt}}(x,k)=\int_{0}^{k} \left[F_{\text{1tt}}(x,k_1)-\frac{k_1}{\sqrt{k_1^2+m_0^2}}F_{\text{1tt,sing}}(x,\omega(k_1))\right]dk_1.
\end{eqnarray}

\begin{figure}[h]
	\begin{center}
		\includegraphics[scale=0.4]{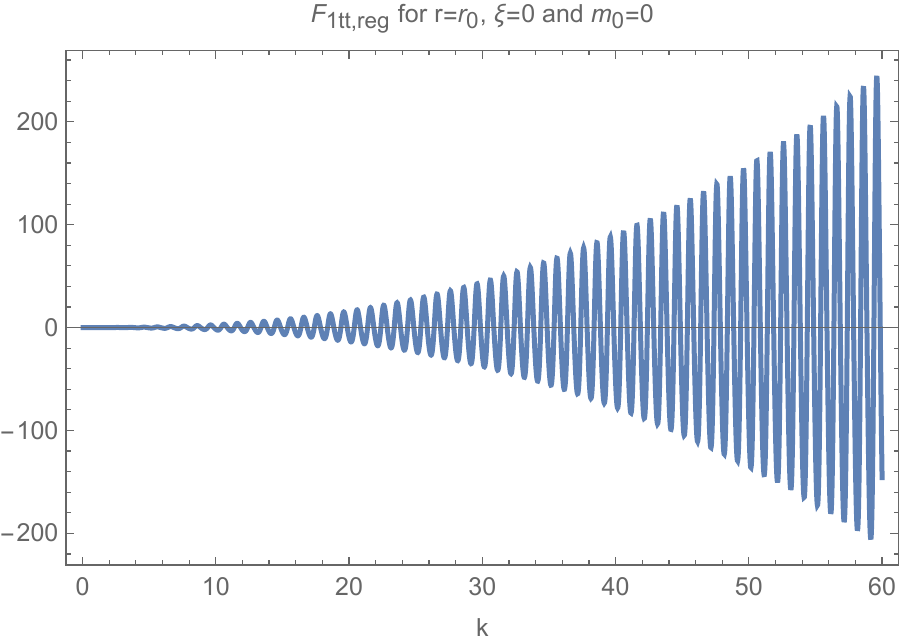}
	\end{center}
	\caption{The curve in this figure displays the value of $F_{\text{1ttreg}}$ for $r=r_0$, $\xi=0$ and $m_0=0$ in the zero-tidal wormhole.}
	\label{figure2}
\end{figure}

\begin{figure}[h]
	\begin{center}
		\includegraphics[scale=0.4]{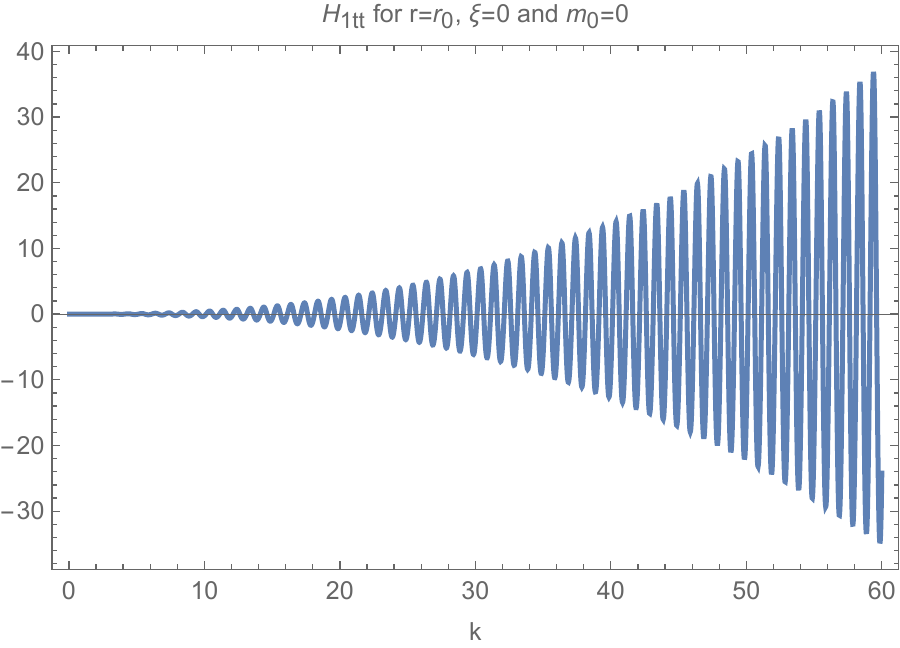}
	\end{center}
	\caption{The curve in this figure displays the value of $H_{\text{1tt}}$ for $r=r_0$, $\xi=0$ and $m_0=0$ in the zero-tidal wormhole.}
	\label{figure3}
\end{figure}

The Fig.\ref{figure2} presents the numerical results of $F_{\text{1ttreg}}$ for $r=r_0$ with $m_0=0$. A comparison between Fig.\ref{figure1} and Fig.\ref{figure2} shows that the divergence of $F_{\text{1ttreg}}$ relative to $F_{\text{1tt}}$ has been substantially alleviated, while $F_{\text{1ttreg}}$ demonstrates strong oscillatory characteristics. The Fig.\ref{figure3} displays the numerical results of $H_{\text{1tt}}$ for $r=r_0$ with $m_0=0$ which is the integral value of the function $F_{\text{1ttreg}}$ with respect to $k$. Comparing Fig.\ref{figure2} and Fig.\ref{figure3}, we observe that the divergence is further weakened. However,‌ the highly oscillatory nature does not vanish after the integration and the result still does not converge. 

\begin{figure}[h]
	\begin{center}
		\includegraphics[scale=0.4]{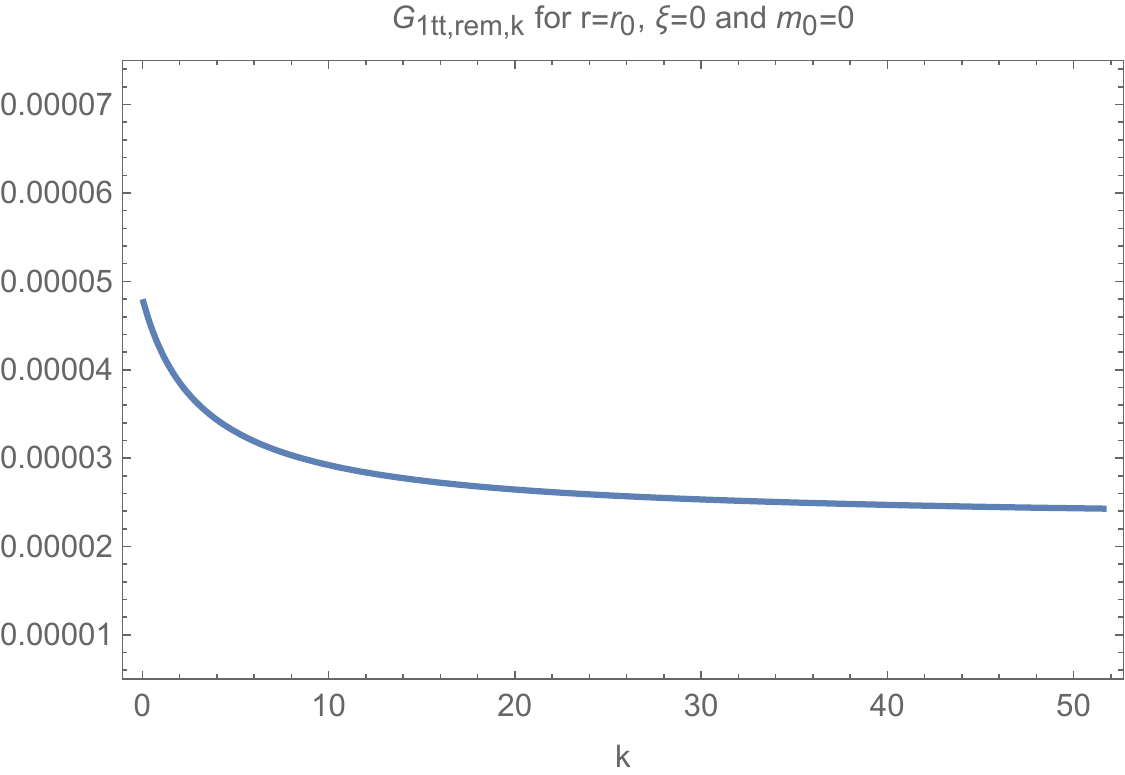}
	\end{center}
	\caption{The curve in this figure displays the value of $G_{\text{1tt,rem,k}}$ for $r=r_0$, $\xi=0$ and $m_0=0$ in the zero-tidal wormhole.}
	\label{figure4}
\end{figure}

The observed oscillations stem from the interchange of the order of limits and integration in the transition from (\ref{g11}) to (\ref{y1}). This operation converts the integral in (\ref{y1}) into a generalized integral. Therefore, we need to employ the self-cancellation method developed in \cite{Levi:2015eea,Levi:2016paz} to handle this generalized integral. We define
\begin{eqnarray}
G_{1uv,\text{rem,k}}(x,k)=\int_{0}^{k} \left[F_{1uv}(x,k)-\frac{k_1}{\sqrt{k_1^2+m_0^2}}F_{1uv,\text{sing}}(x,\omega(k_1))\right]dk_1
-\int_{0}^{m_0}F_{1uv,\text{sing}}(x,\omega(k))d\omega-e_{1uv}(x).
\end{eqnarray}
We have
\begin{eqnarray}
	G_{1uv,\text{rem}}(x)=\lim_{k\rightarrow\infty}G_{1uv,\text{rem,k}}(x,k).
\end{eqnarray}
By applying the self-cancellation method to $H_{1tt}$ and taking into account the second and third terms on the right-hand side of (\ref{y1}), we obtain the numerical results of $G_{\text{1tt,rem,k}}$. Fig.\ref{figure4} presents the numerical results of $G_{\text{1tt,rem,k}}$ for $r=r_0$ with $m=m_0$. From Fig.\ref{figure4}, we observe that $G_{\text{1tt,rem,k}}$ converges to a constant as $k$ increases. By comparing  Fig.\ref{figure1} and Fig.\ref{figure4}, we observe that the numerical magnitude changes from $10^3$ to $10^{-5}$, indicating that high precision must be maintained when solving the mode functions (\ref{mode funtion}).

By repeating the above procedure, we can obtain the numerical results for all $G_{iuv,\text{rem}}$ and $G_{\text{rem}}$. By taking them into (\ref{calculate final}), we obtain the expectation value of the renormalized stress-energy tensor $T_{uv,\text{rem}}$ at the wormhole throat. 

\begin{figure}[h]
	\begin{center}
		\includegraphics[scale=0.4]{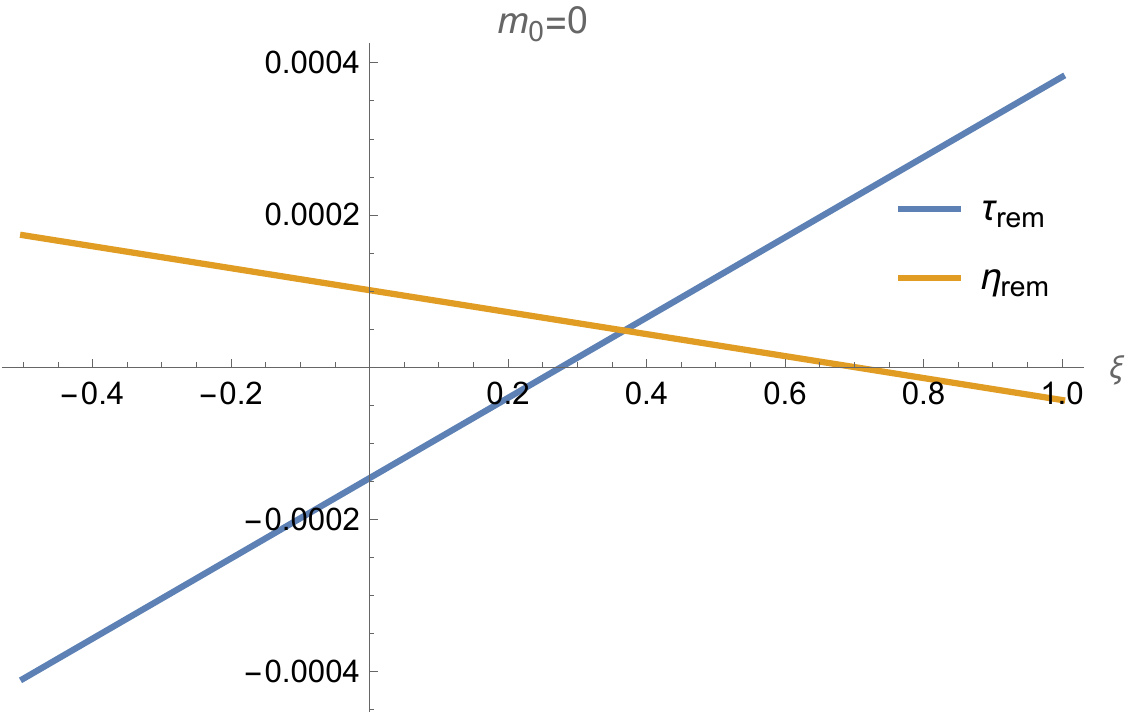}
	\end{center}
	\caption{The curves in this figure display the numerical values of $\tau_{\text{rem}}$ and $\eta_{\text{rem}}$ for $m_0=0$.}
	\label{figure5}
\end{figure}

Now, we can check whether the expectation value of the quantum vacuum stress-energy tensor can satisfy the Morris-Thorne conditions given by (\ref{MT1}) and (\ref{MT2}). We define
\begin{eqnarray}
	\tau_{\text{rem}}=-T_{r_*r_*,\text{rem}}(r_0),\label{QMT1}\\
	\eta_{\text{rem}}=-T_{\text{tt},\text{rem}}(r_0)-T_{r_*r_*,\text{rem}}(r_0).\label{QMT2}
\end{eqnarray}
The Fig.\ref{figure5} presents the numerical results of $\tau_{\text{rem}}$ and $\eta_{\text{rem}}$ for $m_0=0$. From Fig.\ref{figure5}, we observe that there exists a parameter interval for $\xi$ where $\tau_{\text{rem}}>0$ and $\eta_{\text{rem}}>0$. When the coupling constant $\xi$ lies within this interval, the quantum vacuum stress-energy tensor tends to maintain the wormhole structure. When the coupling constant $\xi$ lies outside this interval, the quantum vacuum stress-energy tensor tends to disrupt the wormhole structure. Thus, we find that for massless case, the coupling constant $\xi$ plays opposite roles in wormhole stability across different parameter regions. 

\begin{figure}[h]
	\begin{center}
		\includegraphics[scale=0.4]{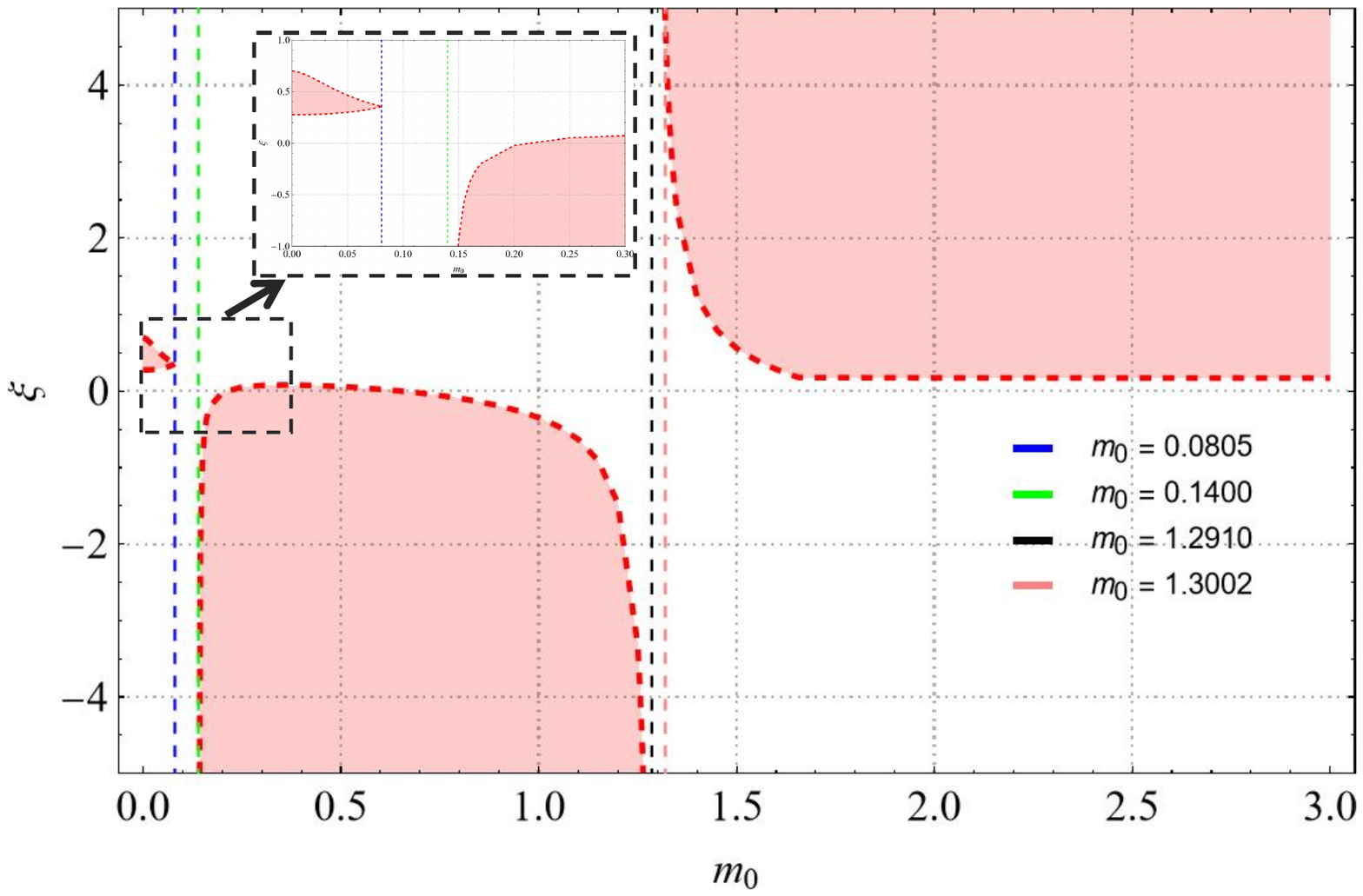}
	\end{center}
	\caption{This figure illustrates how the Morris-Thorne conditions depend on the scalar field mass $m_0$ and the coupling constant $\xi$. The red regions indicate where the Morris-Thorne conditions are satisfied.}
	\label{figure6}
\end{figure}

To systematically investigate the impact of scalar field mass $m_0$ and coupling constant $\xi$ on the stability of the wormhole throat, we perform calculations for various parameter values to explore whether the vacuum renormalized energy-momentum tensor satisfies the Morris-Thorne conditions. In Fig.\ref{figure6}, we explore the Morris-Thorne conditions in the parameter space spanned by the scalar field mass $m_0$ and coupling constant $\xi$. The red regions indicate that the renormalized energy-momentum tensor satisfies the Morris-Thorne conditions, thereby maintaining the wormhole structure. The Fig.\ref{figure6} clearly reveals three disconnected regions satisfying the Morris-Thorne conditions. There exist four critical masses, $m_{1}=0.0805$, $m_{2}=0.1400$, $m_{3}=1.2910$ and $m_{4}=1.3002$. When $m_{0}>m_{4}$, the Morris-Thorne conditions are satisfied only for $\xi>0$, corresponding to the upper-right red region in Figure \ref{figure6}. In this parameter regime, the associated vacuum stress-energy tensor tends to support the wormhole structure. When the scalar field mass $m_0$ lies in the range $m_{\text{2}}<m_0<m_{\text{3}}$, the set of allowed $\xi$ values is overwhelmingly dominated by negative coupling constants, with positive instances constituting a negligible minority. This corresponds to the red area in the bottom-left part of Figure \ref{figure6}. If the scalar field parameters lie in this region, its vacuum stress-energy tensor tends to support the wormhole structure. When $0<m_0<m_{1}$, there is a small parameter window allowing the Morris-Thorne conditions to hold.

It is interesting to note that there exist two intervals $I_{\text{ex1}}=(m_{1},m_{2})$ and $I_{\text{ex2}}=(m_{3},m_{4})$ such that when $m\in I_{\text{ex1}}\cup I_{\text{ex2}}$, no value of the coupling constant $\xi$ can satisfy the Morris-Thorne conditions. Thus, the union $I_{\text{ex1}}\cup I_{\text{ex2}}$ represents a strict no-go regime for wormhole solutions.

\section{Conclusions} \label{section5}
In this paper, we study the renormalized‌ stress-energy tensor of a non-minimally coupled massive scalar field in a zero-tidal wormhole. Within the Hadamard renormalization framework, we numerically calculate the renormalized stress-energy tensor using the pragmatic mode-sum regularization method recently established by Levi and Ori \cite{Levi:2015eea,Levi:2016paz}. By varying the mass $m_0$ and coupling constant $\xi$, we identify the parameter regions in the $m_0$-$\xi$ space where the Morris-Thorne conditions hold. These results are presented in Figure \ref{figure6}. 

‌As shown in Fig.\ref{figure6}, the scalar field mass $m_0$ and coupling constant $\xi$ exert a non-trivial influence on the Morris-Thorne conditions. There exist three disconnected parameter regions that satisfy the Morris-Thorne conditions. For $m_0>m_{4}$, the Morris-Thorne conditions is satisfied only for $\xi>0$. However, for $m_{2}<m_0<m_{3}$, the vast majority of $\xi$ values that satisfy the Morris-Thorne conditions are negative.‌ Furthermore, within the interval $0<m_0<m_{1}$, there exists a parameter region that is significantly smaller than the other two regions and satisfies the Morris-Thorne conditions. When $m\in(m_1,m_2)\cup(m_3,m_4)$, no value of $\xi$ can satisfy the Morris-Thorne conditions. Thus, this set constitutes a ‌no-go regime‌ for traversable wormhole solutions.

Our results demonstrate that the renormalized energy-momentum tensor can satisfy the Morris-Thorne conditions across a broad range of parameters. Consequently, within these parameter regions, the renormalized energy-momentum tensor tends to maintain the wormhole structure. This suggests the possibility of utilizing the quantum vacuum of a scalar field to construct wormholes. On the other hand, we also identify two parameter regions that together constitute a ‌no-go regime‌ for traversable wormhole solutions.

\section{Acknowledgments}
S. J. is supported by the National Natural Science Foundation of China with Grant No. 12275087 and J. J. is supported by the National Natural Science Foundation of China with Grant No. 12205014.


\begin{thebibliography}{99}


\bibitem{Morris:1988cz}
M.~S.~Morris and K.~S.~Thorne,
Am. J. Phys. \textbf{56}, 395-412 (1988)
doi:10.1119/1.15620



\bibitem{Taylor:1996yu}
B.~E.~Taylor, W.~A.~Hiscock and P.~R.~Anderson,
Phys. Rev. D \textbf{55}, 6116-6122 (1997)
doi:10.1103/PhysRevD.55.6116
[arXiv:gr-qc/9608036 [gr-qc]].

\bibitem{Matyjasek:2020cmi}
J.~Matyjasek,
Phys. Rev. D \textbf{102}, no.2, 024082 (2020)
doi:10.1103/PhysRevD.102.024082
[arXiv:2005.07077 [gr-qc]].

\bibitem{Popov:2005qy}
A.~A.~Popov,
Class. Quant. Grav. \textbf{22}, 5223-5230 (2005)
doi:10.1088/0264-9381/22/24/002

\bibitem{Carlson:2010yw}
E.~D.~Carlson, P.~R.~Anderson, A.~Fabbri, S.~Fagnocchi, W.~H.~Hirsch and S.~Klyap,
Phys. Rev. D \textbf{82}, 124070 (2010)
doi:10.1103/PhysRevD.82.124070
[arXiv:1008.1433 [gr-qc]].

\bibitem{Kocuper:2017aap}
E.~Kocuper, J.~Matyjasek and K.~Zwierzchowska,
Phys. Rev. D \textbf{96}, no.10, 104057 (2017)
doi:10.1103/PhysRevD.96.104057
[arXiv:1709.09034 [gr-qc]].

\bibitem{Popov:2001kk}
A.~A.~Popov,
Phys. Rev. D \textbf{64}, 104005 (2001)
doi:10.1103/PhysRevD.64.104005
[arXiv:hep-th/0109166 [hep-th]].

\bibitem{Khusnutdinov:2003ii}
N.~R.~Khusnutdinov,
Phys. Rev. D \textbf{67}, 124020 (2003)
doi:10.1103/PhysRevD.67.124020
[arXiv:hep-th/0304176 [hep-th]].

\bibitem{Bezerra:2010ix}
V.~B.~Bezerra, E.~R.~Bezerra de Mello, N.~R.~Khusnutdinov and S.~V.~Sushkov,
Phys. Rev. D \textbf{81}, 084034 (2010)
doi:10.1103/PhysRevD.81.084034
[arXiv:1003.0344 [gr-qc]].

\bibitem{Schwinger:1951nm}
J.~S.~Schwinger,
Phys. Rev. \textbf{82}, 664-679 (1951)
doi:10.1103/PhysRev.82.664


\bibitem{DeWitt:1964mxt}
B.~S.~DeWitt,
Conf. Proc. C \textbf{630701}, 585-820 (1964)
1964,

\bibitem{Christensen:1976vb}
S.~M.~Christensen,
Phys. Rev. D \textbf{14}, 2490-2501 (1976)
doi:10.1103/PhysRevD.14.2490


\bibitem{Candelas:1984pg}
P.~Candelas and K.~W.~Howard,
Phys. Rev. D \textbf{29}, 1618-1625 (1984)
doi:10.1103/PhysRevD.29.1618

\bibitem{Howard:1984ttx}
K.~W.~Howard,
Phys. Rev. D \textbf{30}, 2532-2547 (1984)
doi:10.1103/PhysRevD.30.2532
\bibitem{Levi:2015eea}
A.~Levi and A.~Ori,
Phys. Rev. D \textbf{91}, 104028 (2015)
doi:10.1103/PhysRevD.91.104028
[arXiv:1503.02810 [gr-qc]].

\bibitem{Levi:2016paz}
A.~Levi,
Phys. Rev. D \textbf{95}, no.2, 025007 (2017)
doi:10.1103/PhysRevD.95.025007
[arXiv:1611.05889 [gr-qc]].

\bibitem{Decanini:2005eg}
Y.~Decanini and A.~Folacci,
Phys. Rev. D \textbf{78}, 044025 (2008)
doi:10.1103/PhysRevD.78.044025
[arXiv:gr-qc/0512118 [gr-qc]].


\bibitem{DeWitt:1975ys}
B.~S.~DeWitt,
Phys. Rept. \textbf{19}, 295-357 (1975)
doi:10.1016/0370-1573(75)90051-4

\bibitem{Radzikowski:1996pa}
M.~J.~Radzikowski,
Commun. Math. Phys. \textbf{179}, 529-553 (1996)
doi:10.1007/BF02100096













\bibitem{Hollands:2019whz}
S.~Hollands, R.~M.~Wald and J.~Zahn,
Class. Quant. Grav. \textbf{37}, no.11, 115009 (2020)
doi:10.1088/1361-6382/ab8052
[arXiv:1912.06047 [gr-qc]].

\bibitem{Kay:1988mu}
B.~S.~Kay and R.~M.~Wald,
Phys. Rept. \textbf{207}, 49-136 (1991)
doi:10.1016/0370-1573(91)90015-E


\bibitem{Wald:1978pj}
R.~M.~Wald,
Phys. Rev. D \textbf{17}, 1477-1484 (1978)
doi:10.1103/PhysRevD.17.1477


\bibitem{Anderson:1994hg}
P.~R.~Anderson, W.~A.~Hiscock and D.~A.~Samuel,
Phys. Rev. D \textbf{51}, 4337-4358 (1995)
doi:10.1103/PhysRevD.51.4337

\bibitem{DeWitt:1960fc}
B.~S.~DeWitt and R.~W.~Brehme,
Annals Phys. \textbf{9}, 220-259 (1960)
doi:10.1016/0003-4916(60)90030-0


\end{thebibliography}
\end{document}